\documentclass[color]{aib}

\usepackage[utf8]{inputenc}
\usepackage{makeidx}  % allows for indexgeneration
\usepackage{cite}
\usepackage{graphicx}
\usepackage{siunitx}
\usepackage{eurosym}
\usepackage{subfig}
\usepackage{xcolor}
\usepackage{url}
\usepackage{textcomp}
\usepackage[hidelinks]{hyperref}

\newcommand{\techreporttitle}{HotBox: Testing Temperature Effects in Sensor Networks\relax}

\begin{document}

\AIBtitle{{\rwthblue \techreporttitle}}
\AIBauthors{Florian Schmidt, Matteo Ceriotti, Niklas Hauser, Klaus Wehrle}
\AIBnumber{2014}{14}
\AIBdate{December 2014}

\makeAIBtitle

\title{\techreporttitle}

\author{Florian Schmidt\inst{1} \and Matteo Ceriotti\inst{2} \and Niklas Hauser\inst{1} \and Klaus Wehrle\inst{1}}

\institute{Communication and Distributed Systems Group, RWTH Aachen University, Germany\\
\email{\{schmidt,wehrle\}@comsys.rwth-aachen.de, niklas.hauser@rwth-aachen.de}
\and
Networked Embedded Systems Group, University of Duisburg--Essen, Germany\\
\email{matteo.ceriotti@uni-due.de}}

\maketitle

%reset footnote counter so later footnotes don't start at 3
\setcounter{footnote}{0}

\begin{abstract}
Low-power wireless networks, especially in outside deployments, are exposed to a wide range of temperatures.
The detrimental effect of high temperatures on communication quality is well known.
To investigate these influences under controlled conditions, we present HotBox, a solution with the following properties:
(1)~It allows exposition of sensor motes to a wide range of temperatures with a high degree of accuracy.
(2)~It supports specifying exact spatial orientation of motes which, if not ensured, interferes with repeatable experiment setups.
(3)~It is reasonably easy to assemble by following the information (code, PCB schematics, hardware list and crafting instructions) available online, facilitating further use of the platforms by other researchers.
After presenting HotBox, we will show its performance and prove its feasibility as evaluation platform by conducting several experiments.
These experiments additionally provide further insight into the influence of temperature effects on communication performance in low-power wireless networks.
\end{abstract}

\section{Introduction}

Low-power wireless networked devices are seeing more and more uses, enabling monitoring of areas both remote and inaccessible, and within our own homes, from any point on Earth.
Depending on the deployment scenario, those devices can be exposed to strongly varying environmental effects.
In recent publications, it has been shown that temperature has a strong effect on communication quality\cite{bannister08hot,boano10transaction,boano13extreme,wennerstrom13extreme,boano14templab}: as temperature rises, communication becomes more challenging, up to an eventual complete breakdown.
While initial results were extracted from real-world outside deployments, further investigation has largely moved to scenarios in which environmental factors are more controllable.
This has led to solutions such as TempLab\cite{boano14templab}, which allows exposing a large number of sensor motes to strictly controlled temperatures.
In this setup, motes are either inserted into a insulation enclosure that provides a heating and cooling element; however, the used hardware makes these boxes somewhat expensive.
Alternatively, budget setups consist of only the mote and an infrared lamp; the lower cost is offset by lacking the capability to cool the device and, more significantly, the absence of any insulation, which allows a greater impact of environmental influences.
Finally, no setup allows to anchor the devices under test in such a way that experiments are reproducible with a given accuracy.

In this technical report, we present HotBox, an alternative solution to exactly control temperature and spatial orientation of sensor nodes.
Furthermore, we exploit this setup to perform an investigation of the impact of temperature on low-power wireless communication and describe the preliminary result of our measurement campaign.
Most of the research on this topic focuses on the effects of temperature on packet error rate (PER) or packet loss rate (PLR), that is, the fraction of sent packets that are received with errors or not received at all.
While we also present results on this level, we additionally take a more detailed look into the erroneously received packets, and investigate the influence of temperature on the distribution of bit errors within a packet.
Overall, this technical report makes three contributions:
(1)~We present ``HotBox'', our system to exactly control temperature and spatial orientation of sensor nodes. HotBox is a budget solution that nevertheless works very well and can easily be used by other researchers.
(2)~We use HotBox to investigate the influence of temperature on bit error distribution patterns within erroneous messages.
(3)~Using the same setup, we present results regarding various packet- and connection-based metrics.

\section{Related Work}

Following the list of contributions we target in this work, we turn our attention now to the current state of the art in testbeds for the study of the impact of temperature, and knowledge about error distributions in low-power wireless networks and the impact of temperature on them.

\paragraph{WSN Testbeds and Temperature Control}
The most commonly used testbeds of low-power devices, e.g., TWIST~\cite{twist} or Indriya~\cite{indriya}, provide indoor infrastructures to analyze protocol and application behavior in realistic settings with increased visibility through the use of wired back-channels.
Being mostly indoor, they are not well suited to analyze the impact of temperature on system performance.
Bannister et al.~\cite{bannister08hot} were one of the first to provide a systematic analysis of the impact of temperature on the performance of the CC2420 radio, the one also employed in our study.
Two radios were connected together via coaxial cable and exposed to temperatures ranging from \SI{25} to \SI{65}{\celsius}.
While this shields the setup from the effect of radio irregularities and environment interference, it also raises the question of how this abstraction influences the results.
Boano et al.~\cite{boano14templab} introduced Temp\-Lab, a testbed allowing fine-grained analysis of temperature impact on wireless sensor networks based on two different types of setups. 
A budget solution (\euro 65) is composed exclusively of a remotely-controlled IR light bulb placed near the device under test.
Alternatively, the additional use of polystyrene foam enclosures, and Peltier elements allows to build small, more accurate temperature chambers at a relatively low cost (\euro 293 per chamber).
In our work, we aim to explore the ground in between such alternatives, and to make available the solution for everybody to reproduce.

\paragraph{Bit Error Distributions}
The study of errors in low-power 802.15.4 wireless communication has mostly stopped at the packet level, with several experimental studies analyzing PLR.
However, there typically is a large fraction of packets that are received with errors and discarded as corrupted.
Only recently, the study of these erroneous receptions has drawn interest and in-depth experimental studies have been performed.
Schmidt et al.\cite{senserr} examined the bit error distributions within corrupted messages in an outdoor testbed comprising 20 TelosB devices.
Most interestingly, the study observed that within 4-bit symbols, the Most Significant Bit (MSB) is significantly less likely to break and that symbols with MSB set to 1 are more likely to break.
This finding was also corroborated by Wennerström et al.~\cite{wennerstrom13extreme}.
The behavior was subsequently explained by Hermans et al.~\cite{hermans14ewsn} with the CC2420's use of an MSK demodulator to receive the signal sent from an O-QPSK modulator.
While the two modulation schemes are sufficiently similar to allow successful demodulation, the translation of code words introduces the observed skew.
All these studies, however, explore the behavior in general, without focusing on quantitative differences among individual devices or analyzing the impact of temperature on the observed behavior.

\paragraph{Impact of Temperature on Packet Errors}
In the aforementioned study by Bannister et al.~\cite{bannister08hot}, the setting with a pair of nodes connected via coaxial cable was used to gather knowledge about the behavior of low-power devices under the impact of temperature.
The results demonstrated a reduction of RSSI with an increase of temperature, more marked with a heated transmitter.
In~\cite{boano10transaction}, this behavior was also identified, again with larger differences when the transmitter was heated than when the receiver was.
In both studies, the correlation between signal power reduction and temperature increase is identified to be caused by the loss of gain in the CC2420 Low Noise Amplifier (LNA).
These studies are further validated by the long-term measurements taken by Wennerström et al.~\cite{wennerstrom13secon} in an outdoor deployment.
Boano et al.~\cite{boano13extreme} conducted a comprehensive study on the effects of temperature on RSSI, Noise Floor, PLR and LQI, further confirming the asymmetry observed by Bannister et al.~\cite{bannister08hot}.
In our work, we further extend the available knowledge by demonstrating deviations from the previously reported behavior, in particular showing a greater impact of an heated receiver on the decrease in link quality.

\section{HotBox: a Budget Solution for Exact Temperature Control}

\begin{figure}[t]
\centering
\includegraphics[width=9.75cm]{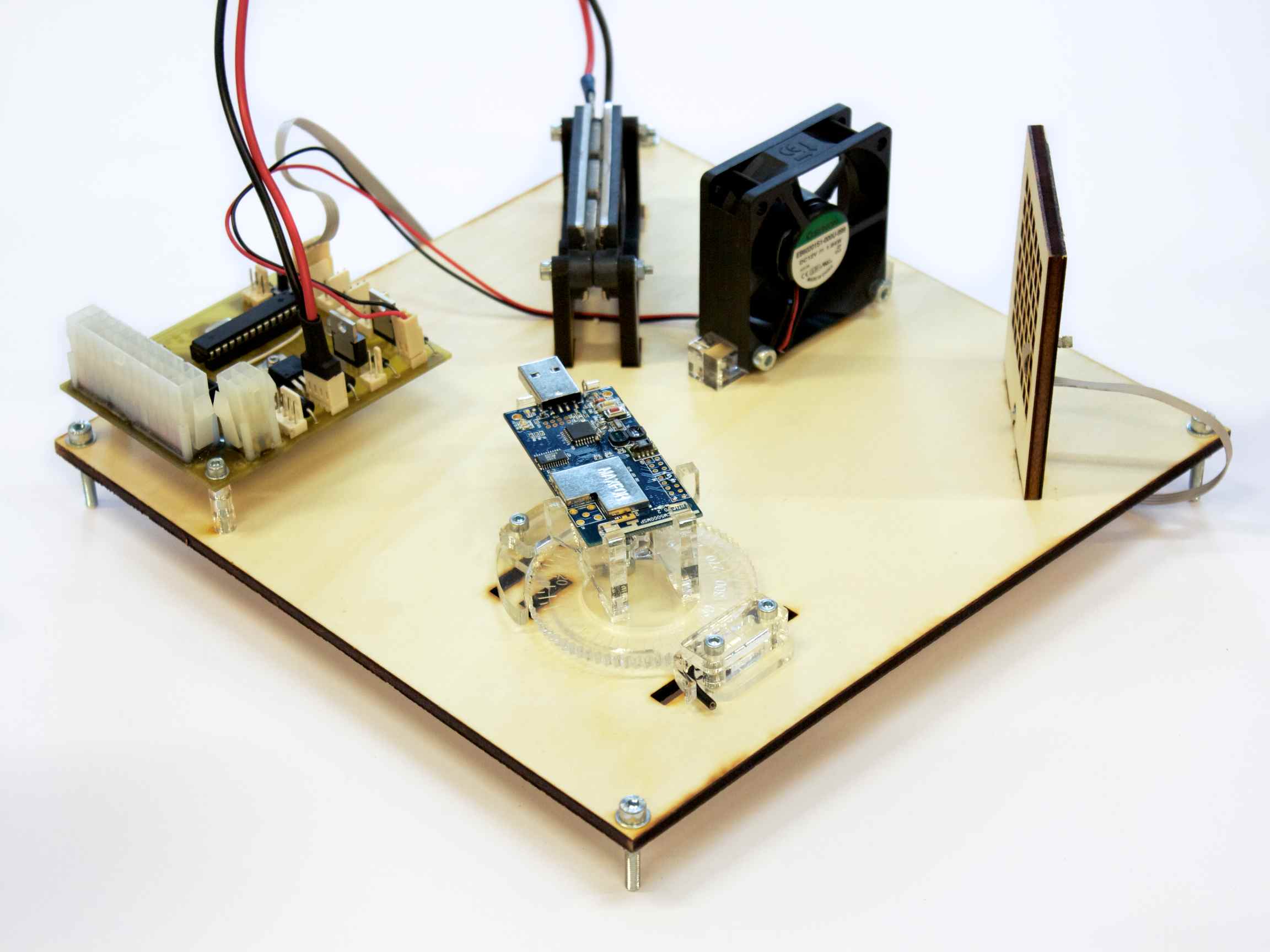}
\caption{HotBox hardware. The control board measures temperature and drives the heating element. The fan and diffusor facilitate even temperature distribution within the enclosure. The mote is fastened to a harness that ensures exact spatial orientation. All parts are mounted on a baseplate, which fits into a polystyrene box (not depicted).}
\label{fig:hotbox}
\end{figure}

One goal of our work was to design a hardware solution to easily control temperature effects on sensor nodes which can be produced relatively quickly and cheaply.
To facilitate this, all hardware elements are off-the-shelf items, while all manufacturing can be done with a soldering iron, a PCB mill, and a laser cutter, which are often available in university settings via the rapidly-spreading FabLab concept.

The core part of the box is the heating element, a \SI{150}{W} ceramic heater which is driven by an ATmega328 controllerwhich in turn is controlled via USB by a central entity that also can be (and in our experiments was) used to monitor and record the experiment results.
Up to five temperature sensors can be distributed within the box to verify and manage homogeneous temperature distribution.
A standard ATX power supply is connected to the control board to provide power to the controller, the heating elements, and the fan.
Since in our experiments, the motes were either powered by batteries or by the USB cables already needed for data exchange, there is currently no provision to power the motes from the control board; however, since the ATX power supply already provides a \SI{5}{V} circuit, the control board could be easily extended to also power the mote.
All elements are mounted to a wooden baseplate, which then is inserted into a polystyrene box of roughly cubic dimensions with a \SI{30}{cm} edge length.
We chose polystyrene for its high thermal insulation capabilities combined with low attenuation of wireless signals\cite{ryan02master} and easy and cheap availability in box form.
The setup as used in our experiments is depicted in Figure~\ref{fig:hotbox}.
All in all, each box costs about \euro 90 in hardware.

\begin{figure}[t]
\centering
\subfloat[Heating from \SI{30}{\celsius} to \SI{90}{\celsius}, then letting it cool back to \SI{30}{\celsius}]{\label{fig:boxperf:speed}\includegraphics[width=0.47\textwidth]{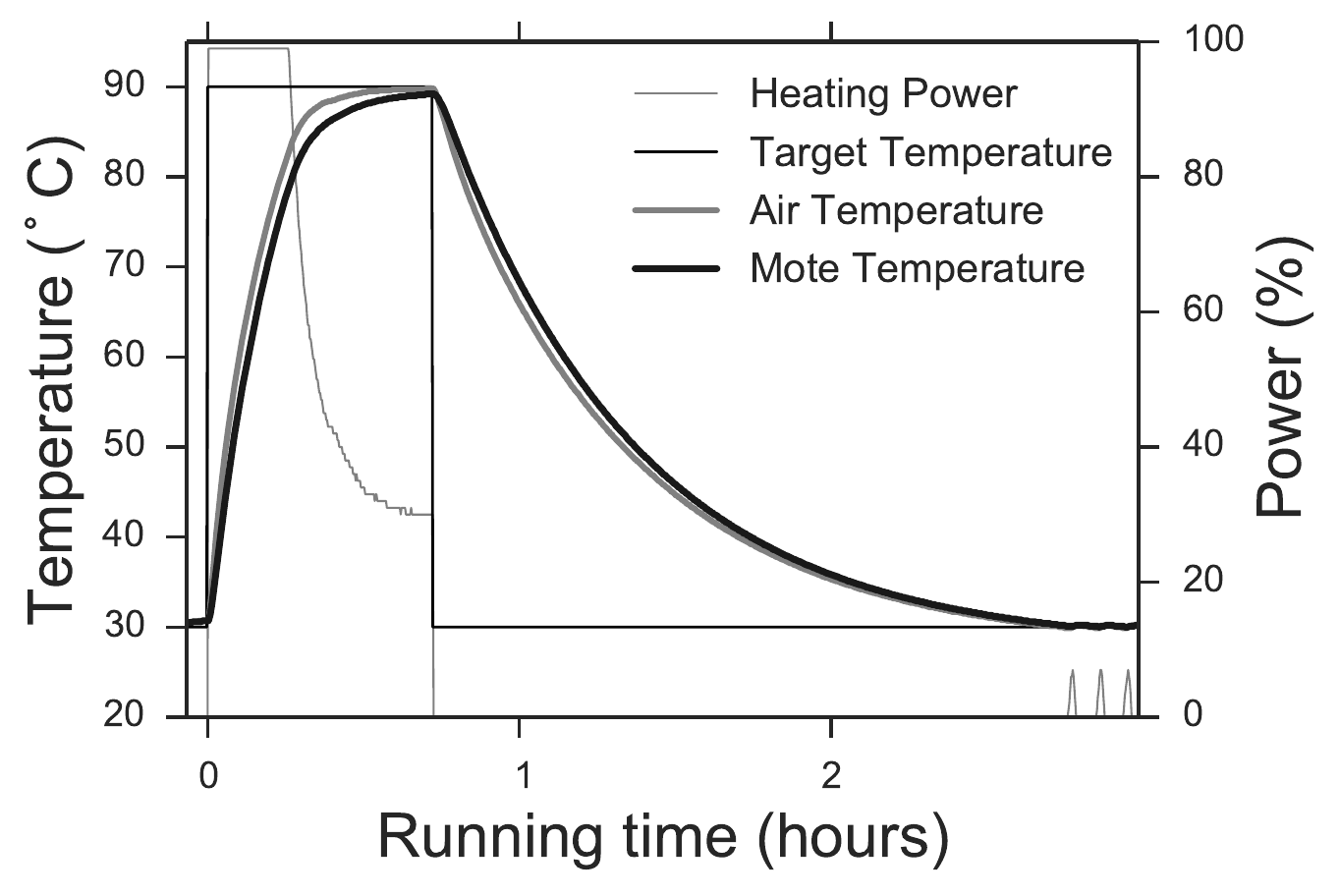}}
\hfill
\subfloat[Heating from \SI{30}{\celsius} to \SI{90}{\celsius} in steps of \SI{5}{\celsius}]{\label{fig:boxperf:accuracy}\includegraphics[width=0.47\textwidth]{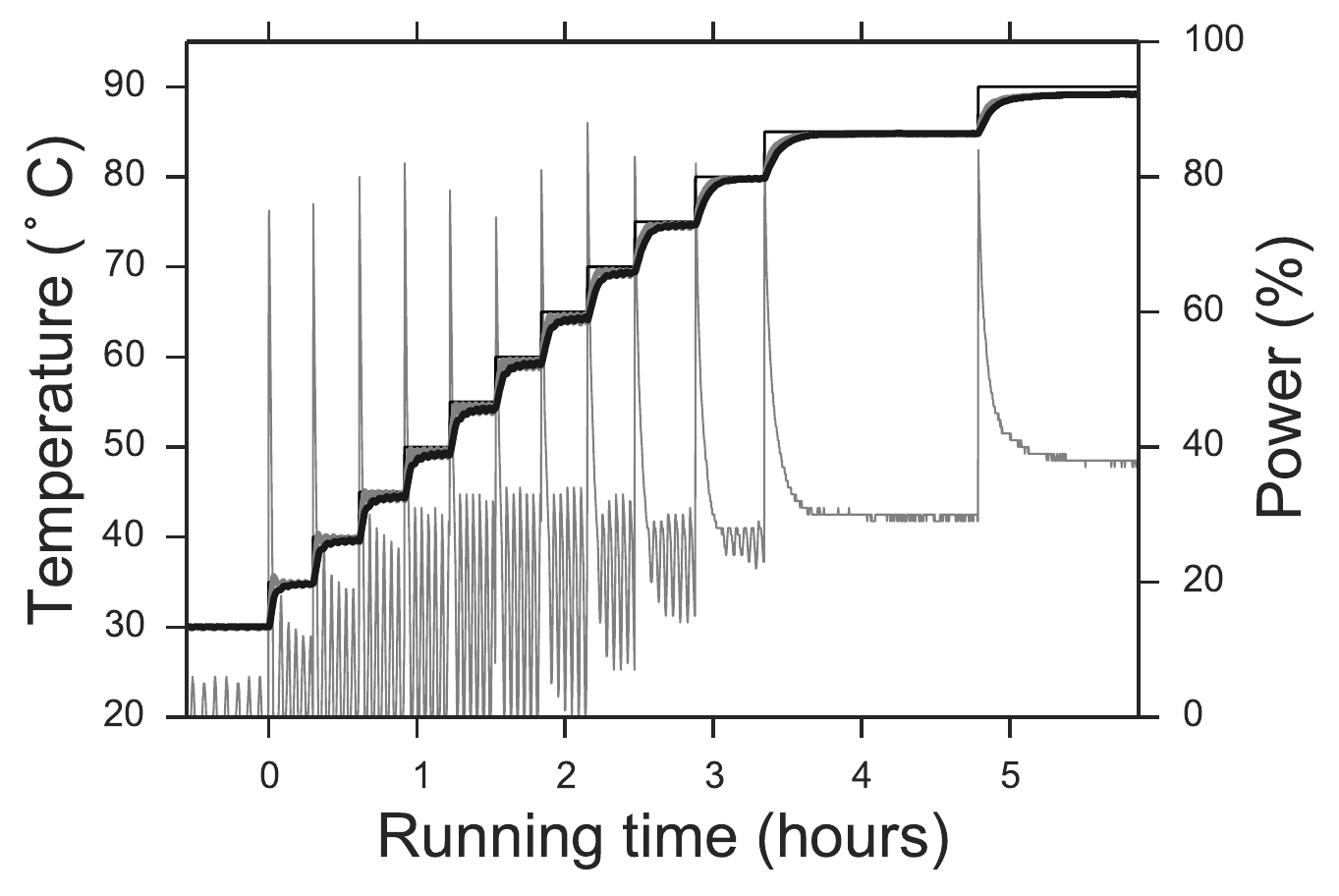}}
\vspace{-6pt}
\caption{HotBox facilitates accurate temperature control. Since the current setup does not include a cooling element, cooling relies on heat exchange with the environment and takes significantly longer than heating. Hence, the heater control loop was designed to not overshoot the target temperature, which is apparent from the asymptotical approach of actual temperature to the target temperature, a result of reduction in power to the heating element as the the two converge.}
\label{fig:boxperf}
\end{figure}

Figure~\ref{fig:boxperf} shows the performance of the box in terms of timeliness and accuracy.
We pre-heated the box to \SI{30}{\celsius} and then increased the target temperature to our allowed maximum of \SI{90}{\celsius}.\footnote{While we tested HotBox with temperatures up to \SI{120}{\celsius}, the point at which the microcontroller stopped functioning, we deliberately limited the maximum temperature to \SI{90}{\celsius} in our experiments to stay within the specifications of the microcontroller and reduce possible fire hazards.}
As can be seen in Figure~\ref{fig:boxperf:speed}, temperature increases almost linearly up to approximately \SI{85}{\celsius} within 20 minutes, then spends another 20 minutes for the last \SI{5}{\celsius}.
After reaching \SI{90}{\celsius}, we switched the target temperature back to \SI{30}{\celsius}.
Since HotBox does not provide any active cooling, the temperature follows a natural cooling pattern with decreasing gradient as the box temperature approaches the environment temperature.
Overall, it takes approximately two hours to cool down from \SI{90}{\celsius} to \SI{30}{\celsius}.
This time can be significantly shortened by removing the polystyrene lid, facilitating ventilation.
Another solution would be to add active cooling to the box.
While we decided against this to keep costs low, HotBox has support to complement the heating unit with a cooling system: the control board already provides additional outlets, while the software would need to be adapted to also drive the cooling element.

During our experiments, we used small temperature steps, so time spent for heating and cooling was secondary to the accuracy of the chosen temperature.
Figure~\ref{fig:boxperf:accuracy} shows the temperature over time when slowly stepping up the temperature in 5-degree steps, similarly to the setup used in the evaluation presented later.
Since the box is only designed for heating and not active cooling, control was set up in a way to prevent overshooting of temperature.
With the TelosB motes used in our experiments, it takes approximately 10--15 minutes to heat them to within \SI{0.5}{\celsius} of the target temperature.
The figure clearly shows that temperature can be controlled to a high degree of accuracy.

The figures also show that mote temperature closely follows air temperature.
Our box setup could therefore also easily be used to accurately control temperature of similar motes which do not provide their own temperature sensor by providing a small ``lag factor'' to account for the slower temperature change within the sensor as opposed to the air.

\begin{figure}[t]
\centering
	\subfloat[Harness with mote on a distance bar.] {
		\includegraphics[width=0.47\textwidth]{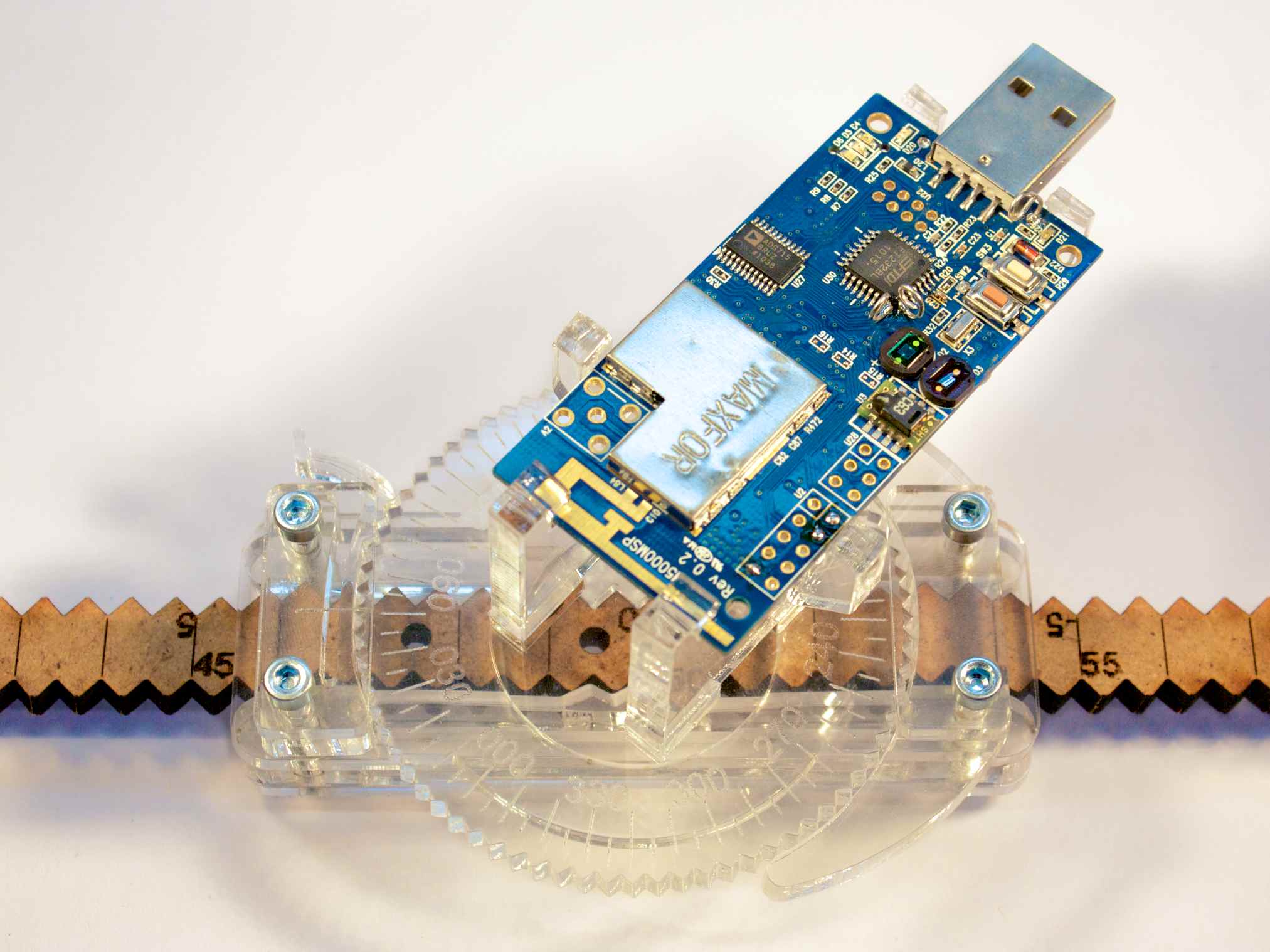}
		\label{fig:harness_picture}
	}
	\subfloat[Vector drawing of the rotation disc.] {
		\includegraphics[trim = 20mm 40mm 140mm 45mm, clip, width=0.5\columnwidth, angle=90, scale=0.75]{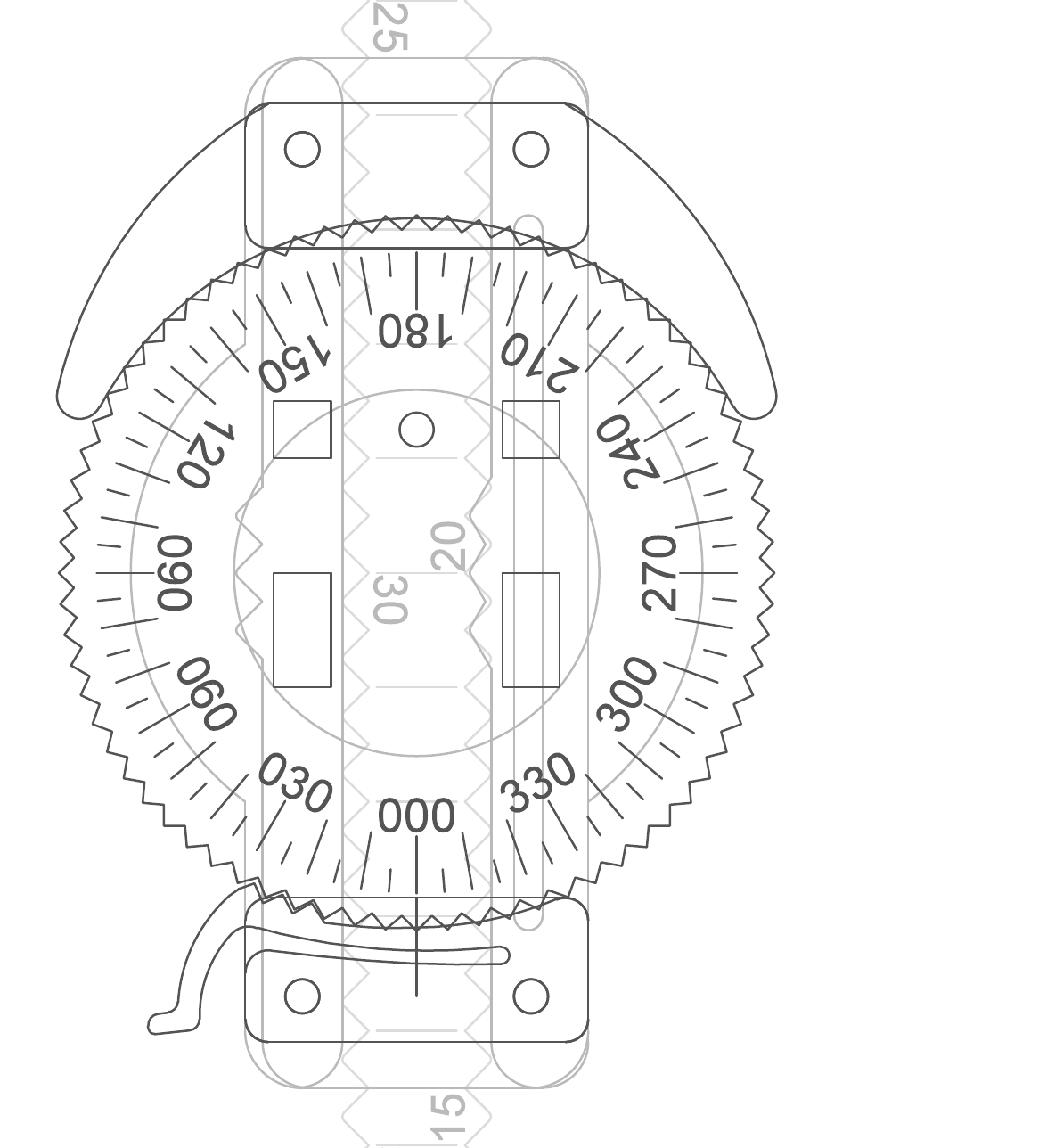}
		\label{fig:harness_vector}
	}
	\caption{The mote harness allows precise positioning of motes. To set orientation, a disc rotates around the center of the PCB antenna at a \SI{5}{\degree} resolution. To set distance (if not fastened to HotBox's baseplate), the harness can slide along distance bars at a \SI{5}{\milli\metre} resolution.}
	\label{fig:mote_harness}
\end{figure}

In order to guarantee that both sender and receiver will not move during experiments, we developed a mote harness that was laser-cut out of acrylic.
The mote snaps into the harness so that the axis of rotation goes through the middle of the PCB antenna which then allows rotation in \SI{5}{\degree} steps as shown in Figure~\ref{fig:mote_harness}.
The reasoning for this is that, although the devices used in our experiments are equipped with a PCB-printed F-type antenna that nominally is omnidirectional, we witnessed strongly varying signal quality depending on the orientation of the nodes.
By using the harness, spatial orientation could be reproduced exactly, supporting repeatability of experiments, especially when nodes are exchanged with each other between experiment runs to investigate potential influences of mote revisions and producers.
We originally developed these harnesses independent of HotBox.
By fastening them to distance bars, exact (to a resolution of \SI{5}{\milli\metre}) and repeatable distance settings were also possible.
While this is still possible, for use with HotBox, the harnesses are instead fastened to the baseplate.
Thus, spatial orientation is still controllable, while distance has to be set by moving the polystyrene boxes.

In summary, HotBox facilitates temperature experiments on sensor motes by exposing them to a highly accurately controllable temperature.
Furthermore, HotBoxes are cheap to produce.
To allow other researchers to conduct temperature experiments using HotBoxes, we made the PCB schematics and control software available online\cite{hotbox-sources}.

\section{Experimental Results}
\label{sec:experiments}

In the following, we will present results from our experimental setups on the influence of temperature on the error distribution of packet communications of sensor nodes.
This presentation serves a dual purpose: first, it shows the practical applicability and usability of HotBox for conducting experiments; second, the results deepen the knowledge about factors such as bit error distributions within frames (see Section~\ref{sec:biterrors}) and various other communication-related metrics (see Section~\ref{sec:packeterrors}) under the influence of temperature.

\subsection{Experimental Setup}

For our experiments, we used TelosB sensor nodes from two different manufacturers as well as production runs and ages.
The TelosB is a widely-used platform that employs a Sensirion SHT11\cite{sensirion} temperature sensor and a CC2420\cite{cc2420} radio chip for communication which implements the IEEE 802.15.4 standard\cite{ieee802154}.
At the physical layer, the standard defines a DSSS (Direct Sequence Spread Spectrum) O-QPSK modulation in the 2.4 GHz ISM band, with a nominal data rate of 250 kbps.

All experiments were conducted in a university room that witnesses little interference from surrounding IEEE 802.11 (WiFi) networks operating in the same band; furthermore, we chose to use channel 26 of the 802.15.4 standard, which is outside the band allocated to 802.11 in Europe.

Experiments used direct (single-hop) connections between links.
Each experiment comprised two boxes with one node each.
Both nodes were connected to a PC via USB; the PC created the packets and sent them via USB to one node; the node would then send the packet via the CC2420 radio.
If the other node received the packet, it forwarded the received version to the PC via USB, which then compared the original and the received version for bit errors.
Otherwise, a timeout would be triggered at the PC to identify the missed reception.
Each run comprised 180\,000 packets.
We exchanged the nodes between experimental runs to account for potential performance differences between production runs and different models of the TelosB nodes.
However, for most investigated metrics, we could not find any noticeable differences in the performance of those nodes. Unless specifically pointed out (cf.\ Section~\ref{sec:biterrors}), the presented results are therefore representative of all investigated node types.

\subsection{Bit Error Distributions}
\label{sec:biterrors}

Previous work\cite{senserr} had analyzed the distribution of bit errors within messages, and found out that some content is inherently more stable than other.
Expanding on this point, we investigated whether changes in temperature change the bit error distribution in packets.
To evaluate this fact, we sent messages with 80 bytes of payload containing a specific pattern between the sensor nodes. Each packet payload was created from a pattern of 0x0000, 0x1111, \dots, 0xFFFF, repeating as necessary to fill the payload.
The reason for this pattern is that the IEEE 802.15.4 standard employs a system in which the packet is split into 4-bit blocks (so-called nibbles) that are then replaced by a 32-bit chipping sequence and sent via DSSS.
Hence, nibbles are the smallest ``atomic'' unit of 802.15.4, and the pattern mentioned above repeats each nibble 4 times before sending the next one.

\begin{figure}[tb]
\centering 
\subfloat[Bit error distribution at \SI{30}{\celsius}]{\label{fig:biterrortemp:30}\includegraphics[width=0.47\textwidth]{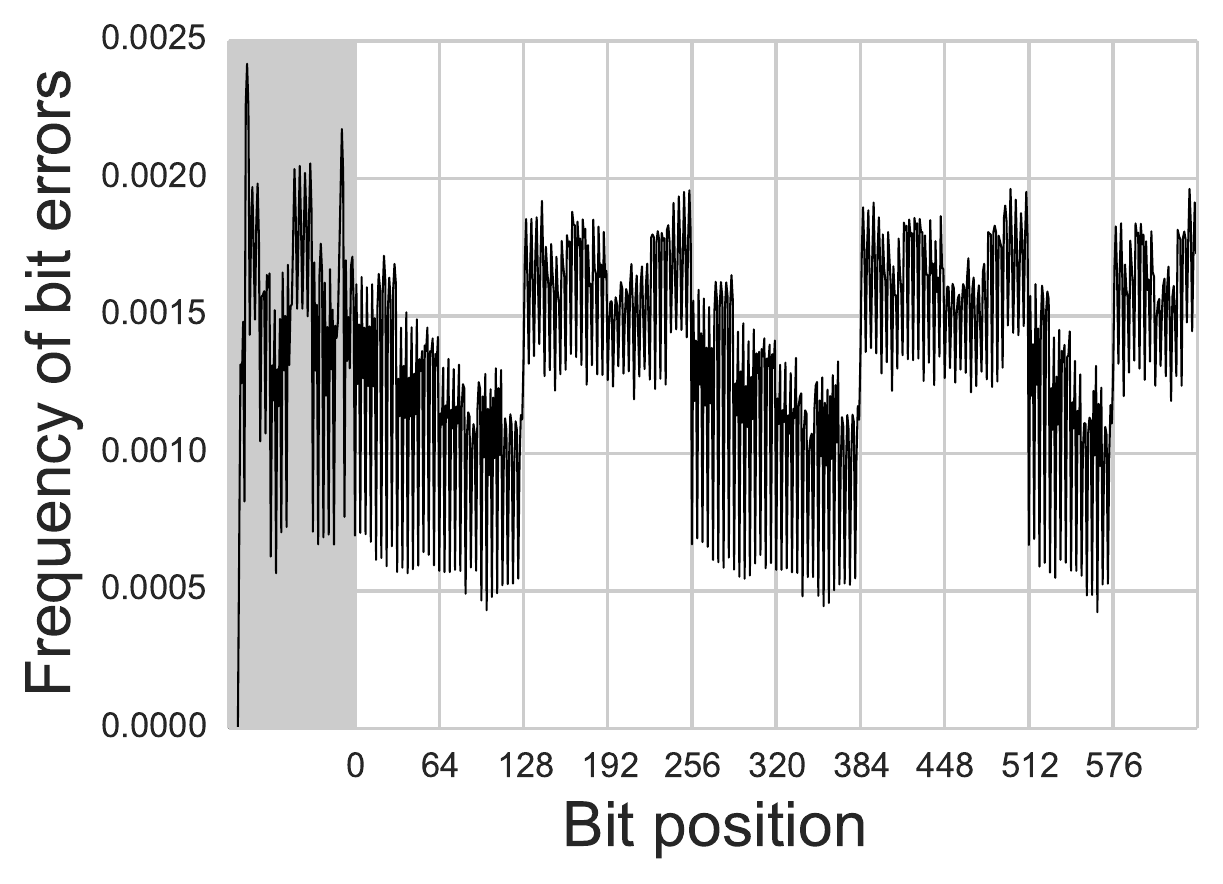}}
\hfill 
\subfloat[Bit error distribution at \SI{70}{\celsius}]{\label{fig:biterrortemp:70}\includegraphics[width=0.47\textwidth]{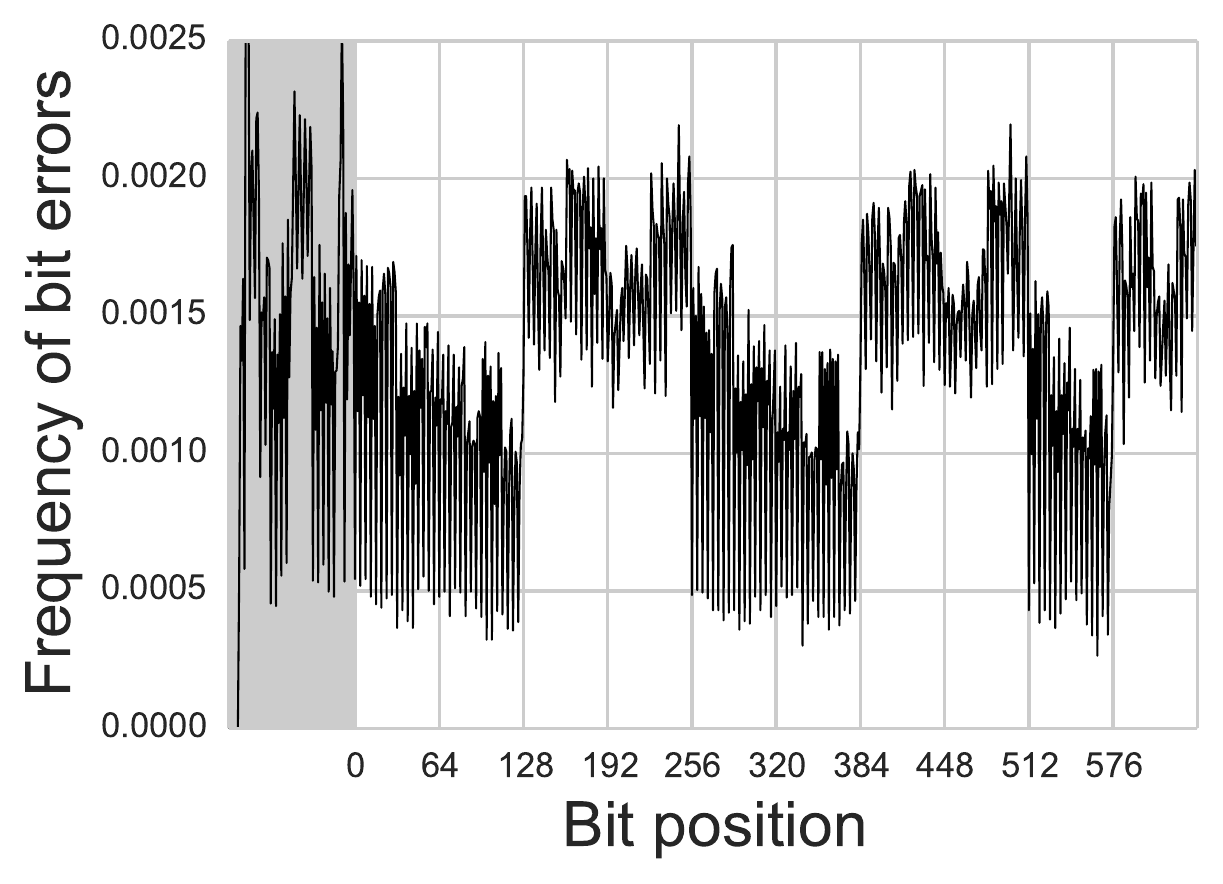}}
\vspace{-6pt}
\caption{When repeatedly sending fixed-content payloads, it is apparent that some bit pattern are more susceptible to errors than others. This behavior does not change significantly with temperature. While more bit errors occur at higher temperatures, the relative bit error frequencies when compared to each other stay very similar. (Gray area denotes the header portion of a frame.)}
\vspace{-6pt}
\label{fig:biterrortemp}
\end{figure}

Previous results\cite{senserr} showed that low-value nibbles with a most-significant bit (MSB) of 0 were more robust to bit errors than those with an MSB of 1.
Since first reporting on this behavior, a strong case has been made\cite{hermans14ewsn} that this is due to the hardware of the CC2420 radio, which during reception uses MSK to demodulate the signal instead of O-QPSK, which is compatible, but explains the uneven error distributions.
In our experiments, we could witness the same error distributions at room temperature as reported on by previous work.
When increasing the temperature, the relative distribution of the pattern stays the same, as can been seen in Figure~\ref{fig:biterrortemp}, where we took an example result between two nodes.
While the absolute number of errors within a message, and therefore the Bit Error Rate (BER) increases, the \textit{relative} distribution of errors between different bits stays roughly the same.

\begin{figure}[tb]
\centering
\subfloat[Low susceptibility]{\includegraphics[width=0.47\textwidth]{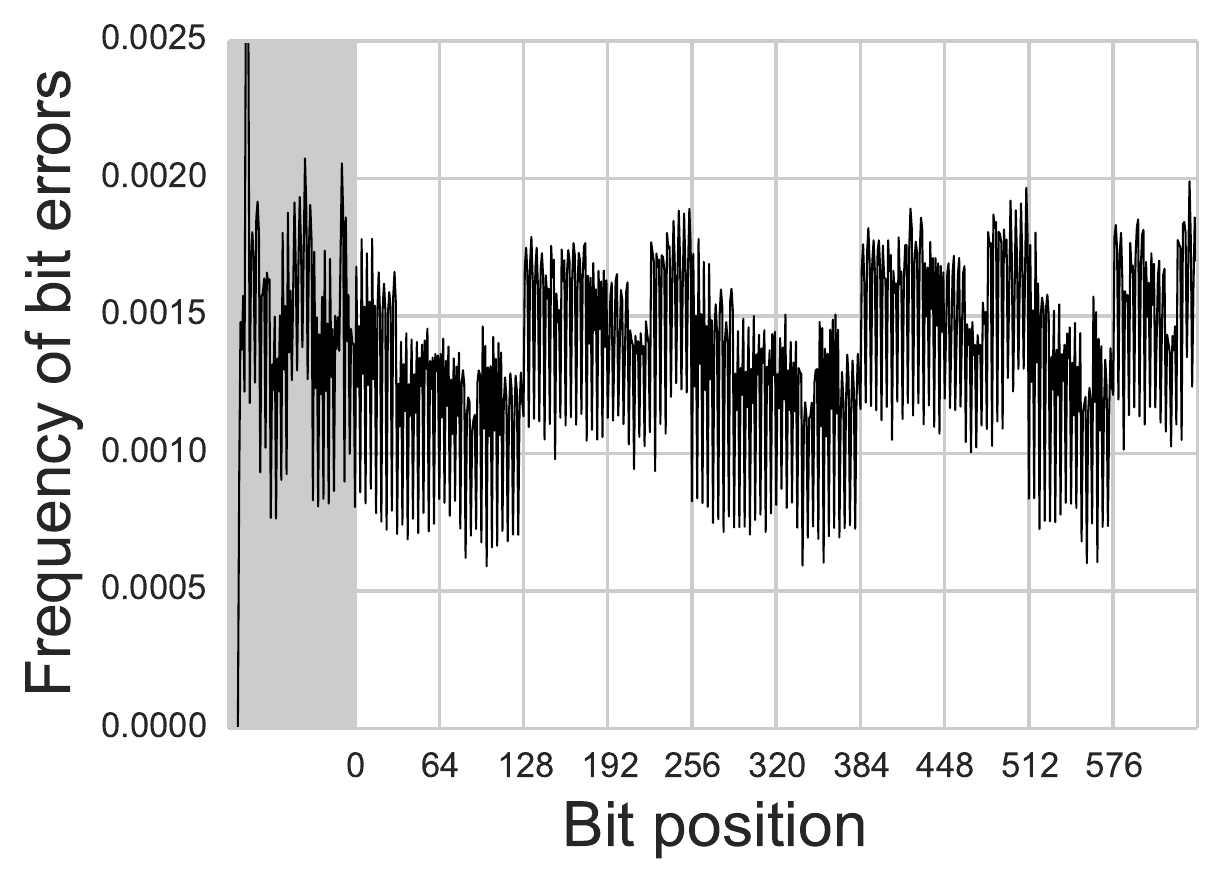}}
\hfill
\subfloat[High susceptibility]{\includegraphics[width=0.47\textwidth]{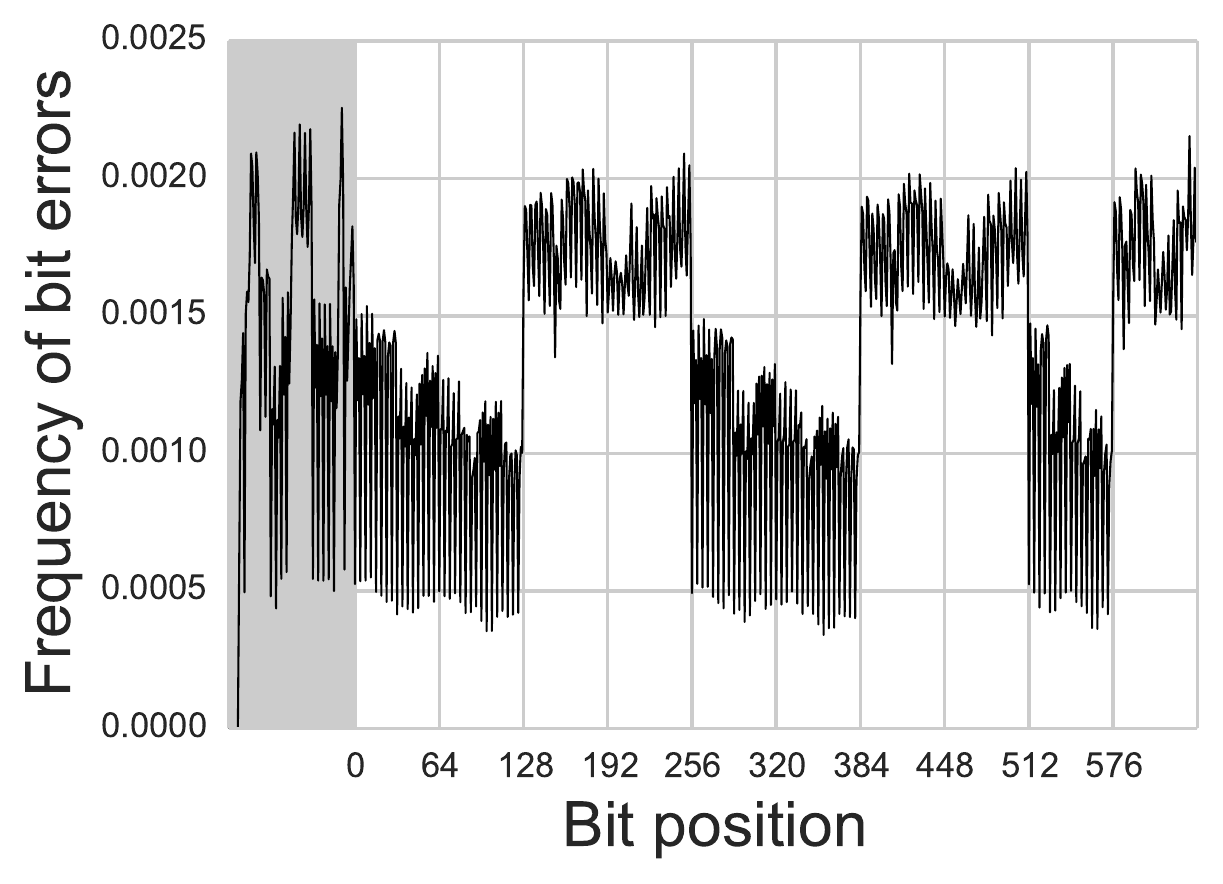}}
\vspace{-6pt}
\caption{Different hardware shows different susceptibility to uneven bit error distributions. While each magnitude of susceptibility is specific to a device and does not change noticeably with temperature or link quality, the differences cannot be traced back to specific production runs or producers; they are effectively random and cannot be predicted without testing. Shown are two examples of susceptibility from nodes at the extreme ends of the spectrum.}
\label{fig:biterrormagnitude}
\vspace{-6pt}
\end{figure}

However, we did notice over the course of our experiments that motes could exhibit this behavior more or less strongly.
While some motes would have 1-MSB nibbles break nearly twice as often as 0-MSB ones, the difference was much more subdued in other cases.
Two examples of results from different motes are given in Figure~\ref{fig:biterrormagnitude}.
These characteristics were specific to a mote and would show up reliably in all tests.
We could not trace back these differences to difference in manufacturer or production run; they seem to rely on effects that are within the production tolerances of the mote and its components.
Without measuring the characteristics, it was effectively impossible to predict the susceptibility of a mote.
Unless a way is found to reliably predict mote characteristics, this significantly complicates ideas of leveraging these imbalances in error distribution to create more robust coding schemes, as suggested in\cite{senserr}.

\begin{figure}
\centering
\subfloat[Transmitter heated]{\label{fig:packeterrors:transmitter}\includegraphics[width=0.47\textwidth]{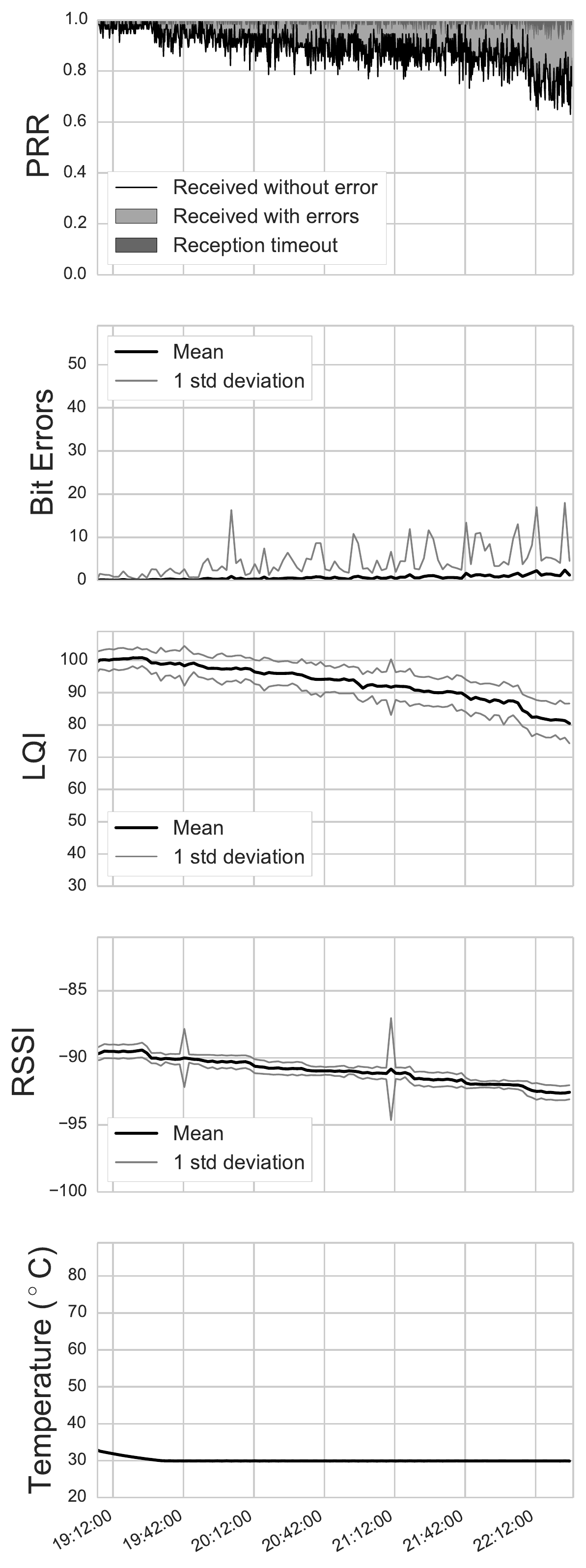}}
\hfill
\subfloat[Receiver heated]{\label{fig:packeterrors:receiver}\includegraphics[width=0.47\textwidth]{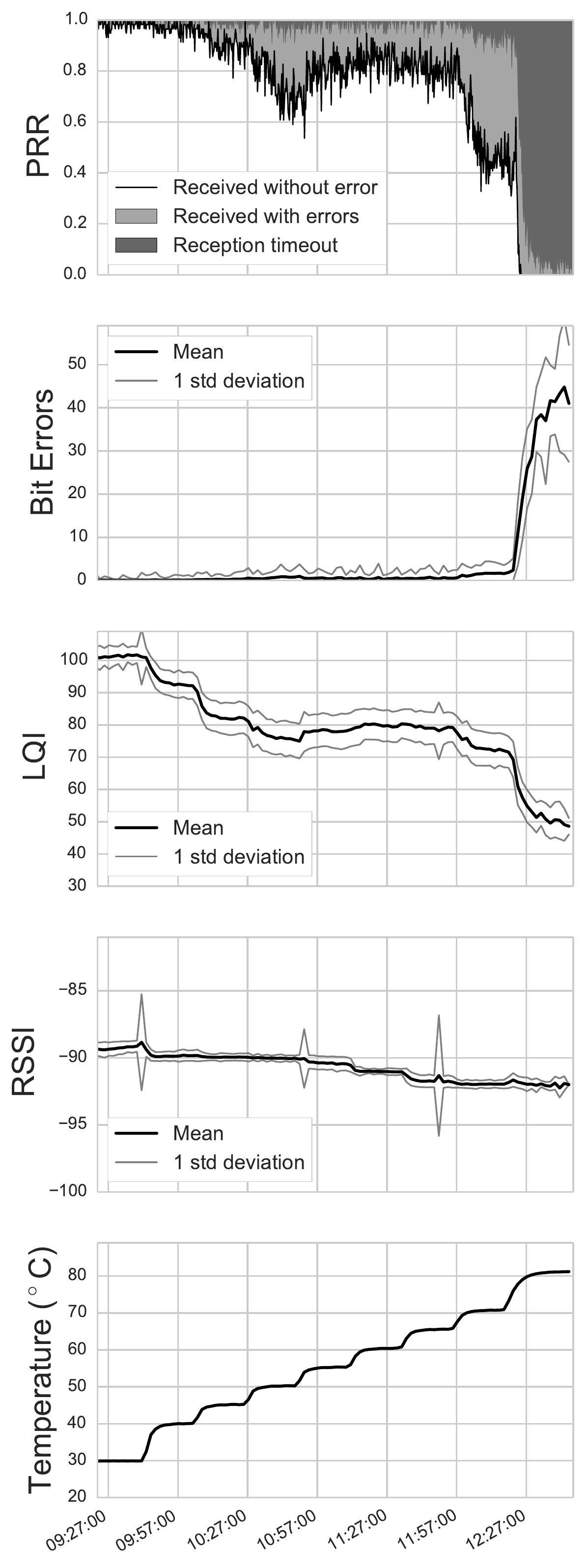}}
\caption{Influence of temperature on several key communication metrics. Note that all results are collected at the receiver, hence temperature in \ref{fig:packeterrors:transmitter} stays stable because only the transmitter is heated (in the same fashion as the receiver in \ref{fig:packeterrors:receiver}). While temperature has a negative effect on all metrics, the effect is much stronger if the receiver is heated, with the exception of RSSI, in which case there are no significant differences between heating the transmitter or the receiver.}
\label{fig:packeterrors}
\end{figure} 

\subsection{Packet Error Dependency on Temperature}
\label{sec:packeterrors}
Previous work\cite{bannister08hot,boano10transaction,boano13extreme,wennerstrom13extreme,boano14templab} has shown a strong influence of temperature on communication quality.
We therefore wanted to check whether we could reproduce these results with our HotBox hardware.
For these experiments, we started with a temperature of \SI{30}{\celsius} and then gradually increased the temperature in 5 or \SI{10}{\celsius} increments, spending 20 minutes at each target temperature before switching to the next higher one.
In these experiments, we distinguished between heating the transmitter and heating the receiver of a connection.
Two nodes took turns sending messages to each other; one box was heated, while the other was kept at a constant temperature of \SI{30}{\celsius}.
Results for each node were saved separately, thereby creating two separate datasets from heating the transmitter and from heating the receiver during a single experiment.
As metrics, we measured the RSSI, the Link Quality Indicator (LQI), and BER as well as packet reception rate (PRR).
We split the latter into three cases: a packet could be received without errors; it could be received, but with errors; or it could be not received at all (completely lost).
The latter case is typically a result of strong corruption within the preamble of the packet, which prevents the receiver from detecting that a packet was sent over the wireless channel.

Figure~\ref{fig:packeterrors} shows a typical result from one of our experiments.
All presented results are values as witnessed by the receiving mote.
As such, the temperature shown in Figure~\ref{fig:packeterrors:receiver} shows the changing temperature settings throughout the experiments.
Conversely, Figure~\ref{fig:packeterrors:transmitter} does not show any changes in the temperature, because it was the transmitter which was heated, while the receiver, whose values are shown, was kept at a constant temperature.

Overall, it can be seen that all metrics are negatively influenced by temperature.
However, the amount as to which they are influenced differs, and heating the transmitter and the receiver has different magnitudes of effect for most metrics.
The only metric that is largely independent of the fact which mote is heated is the RSSI.
All other metrics show a much higher negative influence when the receiver is heated.
Heating the transmitter to \SI{80}{\celsius} still allows communication, albeit with a packet error rate of more than 20\%.
In contrast, communication completely breaks if that temperature is applied at the receiver's side, and even at \SI{70}{\celsius}, PER is much higher at above 50\%.
This is reflected in the BER, which explodes at receiver temperature above \SI{70}{\celsius}.
At the same time, LQI significantly decreases.

Summarizing, our results reinforce the notion of temperature as a significant influence on the communication quality in low-power networks.
However, we were not able to reproduce the results in\cite{boano13extreme}, which showed transmitter heating as the larger influence on quality.
In our experiments, heating the receiver produces a larger impact.

\section{Conclusion}
In this technical report, we first presented a new way to cheaply and accurately control temperature in wireless sensor testbeds.
Our solution is cost-effective, can control temperature to an accuracy of \SI{0.5}{\celsius} as well as orientation to an accuracy of \SI{5}{\degree}, and its schematics and software are available to the public\cite{hotbox-sources}.
We then used this temperature-controlled environment to investigate the effects of temperature on the communication quality of low-power wireless networks, both to validate the usability of HotBox as experimentation platform and to contribute to the current state of knowledge in the area of temperature-related effects on communications in such networks.
We could confirm prior findings by Schmidt et al.\cite{senserr}, and that those effects are independent of temperature.
However, we witnessed differences in the strength of these effects between different sensor nodes, without any apparent correlation with producers or production runs.
At the packet level, we could confirm previous results in that an increase in temperature causes a decrease in communication quality.
However, as opposed to other results, which showed a larger negative effect when the transmitter was heated\cite{bannister08hot,boano10transaction}, we consistently measured a larger influence when the receiver was heated in our experiments.
This difference warrants future, more thorough, investigations to scrutinize the effect of either communication partner on the overall performance.

\bibliographystyle{abbrv}
\bibliography{references}

\begin{thebibliography}{10}

\bibitem{bannister08hot}
K.~Bannister, G.~Giorgetti, and S.~K.~S. Gupta.
\newblock Wireless sensor networking for ``hot'' applications: Effects of
  temperature on signal strength, data collection and localization.
\newblock In {\em Proc.\ HotEmNets}. ACM, June 2008.

\bibitem{boano10transaction}
C.~A. Boano, N.~Tsiftes, T.~Voigt, J.~Brown, and U.~Roedig.
\newblock The impact of temperature on outdoor industrial sensornet
  applications.
\newblock {\em IEEE Transactions on Industrial Informatics}, 6(3):451--459, Aug
  2010.

\bibitem{boano13extreme}
C.~A. Boano, H.~Wennerstr\"{o}m, M.~A. Z{\'u}{\~n}iga, J.~Brown,
  C.~Keppitiyagama, F.~J. Oppermann, U.~Roedig, L.-{\AA}. Nord\'{e}n, T.~Voigt,
  and K.~R\"{o}mer.
\newblock {Hot Packets}: A systematic evaluation of the effect of temperature
  on low power wireless transceivers.
\newblock In {\em Proc.\ ExtremeCom}, pages 7--12. ACM, Aug. 2013.

\bibitem{boano14templab}
C.~A. Boano, M.~Z\'{u}\~{n}iga, J.~Brown, U.~Roedig, C.~Keppitiyagama, and
  K.~R\"{o}mer.
\newblock Templab: A testbed infrastructure to study the impact of temperature
  on wireless sensor networks.
\newblock In {\em Proc.\ IPSN}, pages 95--106. IEEE Press, Apr. 2014.

\bibitem{indriya}
M.~Doddavenkatappa, M.~C. Chan, and A.~L. Ananda.
\newblock Indriya: A low-cost, 3d wireless sensor network testbed.
\newblock In {\em Proc.\ TRIDENTCOM}, pages 302--316. Springer, May 2011.

\bibitem{twist}
V.~Handziski, A.~K\"{o}pke, A.~Willig, and A.~Wolisz.
\newblock Twist: A scalable and reconfigurable testbed for wireless indoor
  experiments with sensor networks.
\newblock In {\em Proc.\ REALMAN}, pages 63--70. ACM, 2006.

\bibitem{hermans14ewsn}
F.~Hermans, H.~Wennerström, L.~McNamara, C.~Rohner, and P.~Gunningberg.
\newblock All is not lost: Understanding and exploiting packet corruption in
  outdoor sensor networks.
\newblock In {\em Proc.\ EWSN}, pages 116--132. Springer International
  Publishing, Feb. 2014.

\bibitem{hotbox-sources}
Hotbox schematics and sources.
\newblock \url{https://github.com/salkinium/hotbox}, 2014.

\bibitem{ieee802154}
{Low-Rate Wireless Personal Area Networks (LR-WPANs)}.
\newblock {IEEE Std.\ 802.15.4}, 2011.

\bibitem{ryan02master}
P.~Ryan.
\newblock Radio frequency propagation differences through various transmissive
  materials.
\newblock Master's thesis, University of North Texas, Dec. 2002.

\bibitem{senserr}
F.~Schmidt, M.~Ceriotti, and K.~Wehrle.
\newblock Bit error distribution and mutation patterns of corrupted packets in
  low-power wireless networks.
\newblock In {\em Proc.\ WiNTECH}, pages 49--56. ACM, Sept. 2013.

\bibitem{sensirion}
{Sensirion}.
\newblock {SHT1x Datasheet}.
\newblock
  \url{http://www.sensirion.com/fileadmin/user_upload/customers/sensirion/Dokumente/Humidity/Sensirion_Humidity_SHT1x_Datasheet_V5.pdf}.

\bibitem{cc2420}
{Texas Instruments}.
\newblock {CC2420 Datasheet}.
\newblock \url{http://focus.ti.com/docs/prod/folders/print/cc2420.html}.

\bibitem{wennerstrom13secon}
H.~Wennerstr\"{o}m, F.~Hermans, O.~Rensfelt, C.~Rohner, and L.-{\AA}.
  Nord\'{e}n.
\newblock A long-term study of correlations between meteorological conditions
  and 802.15.4 link performance.
\newblock In {\em Proc.\ SECON}, pages 221--229, June 2013.

\bibitem{wennerstrom13extreme}
H.~Wennerstr\"{o}m, L.~McNamara, C.~Rohner, and L.-{\AA}. Nord\'{e}n.
\newblock Transmission errors in a sensor network at the edge of the world.
\newblock In {\em Proc.\ ExtremeCom}. ACM, Aug. 2013.

\end{thebibliography}

\cleardoublepage
\section*{Aachener Informatik-Berichte}
\newfont{\sss}{cmr10 scaled 1000}
\newfont{\bbb}{cmbx10 scaled 1000}
\sss

{\bbb This list contains all technical reports published
  during the past three years.
  A complete list of reports dating back to 1987 is available from:
\begin{center}
  \url{http://aib.informatik.rwth-aachen.de/}
\end{center}
  To obtain copies please consult the above URL or send your request
  to:
\begin{center}
  Informatik-Bibliothek, RWTH Aachen, Ahornstr.~55, 52056 Aachen,\\
  Email: \email{biblio@informatik.rwth-aachen.de }
\end{center}}\bigskip

\begin{longtable}{lp{11cm}}

2011-01 $^\ast$ &Fachgruppe Informatik:      Jahresbericht 2011\\
2011-02 & Marc Brockschmidt, Carsten Otto, J\"{u}rgen Giesl:         Modular Termination Proofs of Recursive Java Bytecode Programs by Term Rewriting\\
2011-03 & Lars Noschinski, Fabian Emmes, J\"{u}rgen Giesl:         A Dependency Pair Framework for Innermost Complexity Analysis of Term Rewrite Systems\\
2011-04 & Christina Jansen, Jonathan Heinen, Joost-Pieter Katoen, Thomas Noll:         A Local Greibach Normal Form for Hyperedge Replacement Grammars\\
2011-06 & Johannes Lotz, Klaus Leppkes, and Uwe Naumann:         dco/c++ - Derivative Code by Overloading in C++\\
2011-07 & Shahar Maoz, Jan Oliver Ringert, Bernhard Rumpe:         An Operational Semantics for Activity Diagrams using SMV\\
2011-08 & Thomas Str\"{o}der, Fabian Emmes, Peter Schneider-Kamp, J\"{u}rgen Giesl, Carsten Fuhs:         A Linear Operational Semantics for Termination and Complexity Analysis of ISO Prolog\\
2011-09 & Markus Beckers, Johannes Lotz, Viktor Mosenkis, Uwe Naumann (Editors):         Fifth SIAM Workshop on Combinatorial Scientific Computing\\
2011-10 & Markus Beckers, Viktor Mosenkis, Michael Maier, Uwe Naumann:         Adjoint Subgradient Calculation for McCormick Relaxations\\
2011-11 & Nils Jansen, Erika \'{A}brah\'{a}m, Jens Katelaan, Ralf Wimmer, Joost-Pieter Katoen, Bernd Becker:         Hierarchical Counterexamples for Discrete-Time Markov Chains\\
2011-12 & Ingo Felscher, Wolfgang Thomas:         On Compositional Failure Detection in Structured Transition Systems\\
2011-13 & Michael F\"{o}rster, Uwe Naumann, Jean Utke:         Toward Adjoint OpenMP\\
2011-14 & Daniel Neider, Roman Rabinovich, Martin Zimmermann:         Solving Muller Games via Safety Games\\
2011-16 & Niloofar Safiran, Uwe Naumann:         Toward Adjoint OpenFOAM\\
2011-17 & Carsten Fuhs:         SAT Encodings: From Constraint-Based Termination Analysis to Circuit Synthesis
\\
2011-18 & Kamal Barakat:         Introducing Timers to pi-Calculus\\
2011-19 & Marc Brockschmidt, Thomas Str\"{o}der, Carsten Otto, J\"{u}rgen Giesl:         Automated Detection of Non-Termination and NullPointerExceptions for Java Bytecode\\
2011-24 & Callum Corbett, Uwe Naumann, Alexander Mitsos:         Demonstration of a Branch-and-Bound Algorithm for Global Optimization using McCormick Relaxations\\
2011-25 & Callum Corbett, Michael Maier, Markus Beckers, Uwe Naumann, Amin Ghobeity, Alexander Mitsos:         Compiler-Generated Subgradient Code for McCormick Relaxations\\
2011-26 & Hongfei Fu:         The Complexity of Deciding a Behavioural Pseudometric on Probabilistic Automata\\
2012-01 & Fachgruppe Informatik:         Annual Report 2012\\
2012-02 & Thomas Heer:         Controlling Development Processes\\
2012-03 & Arne Haber, Jan Oliver Ringert, Bernhard Rumpe:         MontiArc - Architectural Modeling of Interactive Distributed and Cyber-Physical Systems\\
2012-04 & Marcus Gelderie:         Strategy Machines and their Complexity\\
2012-05 & Thomas Str\"{o}der, Fabian Emmes, J\"{u}rgen Giesl, Peter Schneider-Kamp, and Carsten Fuhs:         Automated Complexity Analysis for Prolog by Term Rewriting\\
2012-06 & Marc Brockschmidt, Richard Musiol, Carsten Otto, J\"{u}rgen Giesl:         Automated Termination Proofs for Java Programs with Cyclic Data\\
2012-07 & Andr\'{e} Egners, Bj\"{o}rn Marschollek, and Ulrike Meyer:         Hackers in Your Pocket: A Survey of Smartphone Security Across Platforms\\
2012-08 & Hongfei Fu:         Computing Game Metrics on Markov Decision Processes\\
2012-09 & Dennis Guck, Tingting Han, Joost-Pieter Katoen, and Martin R. Neuh\"{a}u\ss{}er:         Quantitative Timed Analysis of Interactive Markov Chains\\
2012-10 & Uwe Naumann and Johannes Lotz:         Algorithmic Differentiation of Numerical Methods: Tangent-Linear and Adjoint Direct Solvers for Systems of Linear Equations\\
2012-12 & J\"{u}rgen Giesl, Thomas Str\"{o}der, Peter Schneider-Kamp, Fabian Emmes, and Carsten Fuhs:         Symbolic Evaluation Graphs and Term Rewriting --- A General Methodology for Analyzing Logic Programs\\
2012-15 & Uwe Naumann, Johannes Lotz, Klaus Leppkes, and Markus Towara:         Algorithmic Differentiation of Numerical Methods: Tangent-Linear and Adjoint Solvers for Systems of Nonlinear Equations\\
2012-16 & Georg Neugebauer and Ulrike Meyer:         SMC-MuSe: A Framework for Secure Multi-Party Computation on MultiSets\\
2012-17 & Viet Yen Nguyen:         Trustworthy Spacecraft Design Using Formal Methods\\
2013-01 $^\ast$ &Fachgruppe Informatik:      Annual Report 2013\\
2013-02 & Michael Reke:         Modellbasierte Entwicklung automobiler Steuerungssysteme in Klein- und mittelst\"{a}ndischen Unternehmen\\
2013-03 & Markus Towara and Uwe Naumann:         A Discrete Adjoint Model for OpenFOAM\\
2013-04 & Max Sagebaum, Nicolas R. Gauger, Uwe Naumann, Johannes Lotz, and Klaus Leppkes:         Algorithmic Differentiation of a Complex C++ Code with Underlying Libraries\\
2013-05 & Andreas Rausch and Marc Sihling:         Software \& Systems Engineering Essentials 2013\\
2013-06 & Marc Brockschmidt, Byron Cook, and Carsten Fuhs:         Better termination proving through cooperation\\
2013-07 & Andr\'{e} Stollenwerk:         Ein modellbasiertes Sicherheitskonzept f\"{u}r die extrakorporale Lungenunterst\"{u}tzung\\
2013-08 & Sebastian Junges, Ulrich Loup, Florian Corzilius and Erika \'{A}brah\'{a}m:         On Gr\"{o}bner Bases in the Context of Satisfiability-Modulo-Theories Solving over the Real Numbers\\
2013-10 & Joost-Pieter Katoen, Thomas Noll, Thomas Santen, Dirk Seifert, and Hao Wu:         Performance Analysis of Computing Servers using Stochastic Petri Nets and Markov Automata\\
2013-12 & Marc Brockschmidt, Fabian Emmes, Stephan Falke, Carsten Fuhs, and J\"{u}rgen Giesl:         Alternating Runtime and Size Complexity Analysis of Integer Programs\\
2013-13 & Michael Eggert, Roger H\"{a}u\ss{}ling, Martin Henze, Lars Hermerschmidt, Ren\'{e} Hummen, Daniel Kerpen, Antonio Navarro P\'{e}rez, Bernhard Rumpe, Dirk Thi\ss{}en, and Klaus Wehrle:         SensorCloud: Towards the Interdisciplinary Development of a Trustworthy Platform for Globally Interconnected Sensors and Actuators\\
2013-14 & J\"{o}rg Brauer:         Automatic Abstraction for Bit-Vectors using Decision Procedures\\
2013-19 & Florian Schmidt, David Orlea, and Klaus Wehrle:         Support for error tolerance in the Real-Time Transport Protocol\\
2013-20 & Jacob Palczynski:         Time-Continuous Behaviour Comparison Based on Abstract Models\\
2014-01 $^\ast$ &Fachgruppe Informatik:      Annual Report 2014\\
2014-02 & Daniel Merschen:         Integration und Analyse von Artefakten in der modellbasierten Entwicklung eingebetteter Software\\
2014-03 & Uwe Naumann, Klaus Leppkes, and Johannes Lotz:         dco/c++ User Guide\\
2014-04 & Namit Chaturvedi:         Languages of Infinite Traces and Deterministic Asynchronous Automata\\
2014-05 & Thomas Str\"{o}der, J\"{u}rgen Giesl, Marc Brockschmidt, Florian Frohn, Carsten Fuhs, Jera Hensel, and Peter Schneider-Kamp:         Automated Termination Analysis for Programs with Pointer Arithmetic\\
2014-06 & Esther Horbert, Germ\'{a}n Mart\'{i}n Garc\'{i}a, Simone Frintrop, and Bastian Leibe:         Sequence Level Salient Object Proposals for Generic Object Detection in Video\\
2014-07 & Niloofar Safiran, Johannes Lotz, and Uwe Naumann:         Algorithmic Differentiation of Numerical Methods: Second-Order Tangent and Adjoint Solvers for Systems of Parametrized Nonlinear Equations\\
2014-08 & Christina Jansen, Florian G\"{o}be, and Thomas Noll:         Generating Inductive Predicates for Symbolic Execution of Pointer-Manipulating Programs\\
2014-09 & Thomas Str\"{o}der and Terrance Swift (Editors):         Proceedings of the International Joint Workshop on Implementation of Constraint and Logic Programming Systems and Logic-based Methods in Programming Environments 2014\\

\end{longtable}
\bigskip

\noindent
{\small $^\ast$ These reports are only available as a printed version.\\
  Please contact \email{biblio@informatik.rwth-aachen.de} to obtain
  copies.}

\end{document}